
\headline={\ifnum\pageno=1\firstheadline\else
\ifodd\pageno\rightheadline \else\leftheadline\fi\fi}
\def\firstheadline{\hfil}
\def\rightheadline{\hfil}
\def\leftheadline{\hfil}
        \footline={\ifnum\pageno=1\firstfootline\else\otherfootline\fi}
\def\firstfootline{\rm\hss\folio\hss}
\def\otherfootline{\hfil}
\font\tenbf=cmbx10
\font\tenrm=cmr10
\font\tenit=cmti10
\font\elevenbf=cmbx10 scaled\magstep 1
\font\elevenrm=cmr10 scaled\magstep 1
\font\elevenit=cmti10 scaled\magstep 1

\nopagenumbers
\line{\hfil }
\vglue 1cm
\hsize=6.0truein
\vsize=8.5truein
\parindent=3pc
\baselineskip=10pt
\centerline{\tenbf GLUEBALLS AND HYBRID MESONS}
\vglue 1.0cm
\centerline{\tenrm CHRIS MICHAEL}
\baselineskip=13pt
\centerline{\tenit D. A. M. T. P., University of Liverpool}
\baselineskip=12pt
\centerline{\tenit Liverpool, L69 3BX, UK.}
\vglue 0.8cm
\centerline{\tenrm ABSTRACT}
\vglue 0.3cm
{\rightskip=3pc
 \leftskip=3pc
 \tenrm\baselineskip=12pt
 \noindent
We discuss states in the meson  spectrum which have explicit
gluonic components. Glueballs (with no valence quarks) and hybrid mesons
(with valence quarks) are both reviewed.  We present in some detail
lattice simulation results.
\vglue 0.8cm }
\line{\elevenbf 1. Introduction \hfil}
\vglue 0.4cm
\baselineskip=14pt
\elevenrm
{}From the early days of QCD, it was clear that the gluonic degrees of
freedom would give rise to some modifications to the simple quark model
of the low-lying hadron spectrum. However, it is now clear
 that there is very little direct
 experimental evidence for these
extra degrees of freedom. This can be understood in a rough way as
arising from the relatively large (circa 1 GeV) energy associated with
the gluonic excitations. Thus the low-lying states do not have much
admixture of gluonic degrees of freedom while those putative states
with gluonic excitations are at larger energies where the spectrum is
dense and less amenable to simple classification.

 In order to set the scene for these
gluonic excitations, it is worthwhile to summarise briefly the
simple constituent quark model. This model of massive quark and
antiquark bound by a potential is only justified theoretically
 for $b\bar{b}$
and to a lesser extent $c\bar{c}$, but it is still useful guide for
light quark states.
The mesonic states that can be
made from $q\bar{q}$ with orbital angular momentum $L$
have $J^{PC}$ values of
\smallskip

\halign{\hskip 0.5in # \hfil & \qquad $#$ \hfil \cr
$L=0$ & 0^{-+}, 1^{--}   \cr
$L=1$ & 1^{+-}, 0^{++}, 1^{++}, 2^{++}  \cr
$L=2$ & 2^{-+}, 1^{--}, 2^{--}, 3^{--}. \cr
}

\smallskip
\noindent Since the gluon can introduce no flavour quantum numbers, the
$J^{PC}$ assignments will be of importance. Of special interest
 in the following will be the absence of
certain $J^{PC}$ values in the above list. These states are known
as spin-exotic and include  $0^{--}, 0^{+-}, 1^{-+}, 2^{+-} $.

In this review, we first discuss the nature of the gluonic excitations
and of the hadrons containing them. We review simple models such as the
MIT bag model and strong-coupling lattice gauge theory. This will allow
a discussion of glueballs.
We then present lattice simulation results
for glueballs which are model-independent. A discussion of the
experimental situation then follows.

As a special case, because it allows a semi-analytic analysis,
we consider glueballs in a small periodic box:
 the femto universe. The  theoretical understanding
of this domain is rather complete and we summarise the
current situation.

We then turn to hybrid mesons: and review bag model, flux tube
and strong-coupling
models before presenting lattice simulation results. Again we discuss
the experimental situation.

Other recent theoretical reviews$^{1,2}$ of gluonic excitations in hadrons
cover many of the topics discussed here but
concentrate rather less on recent lattice results than this review.

\vskip 0.6cm
\line{\elevenbf 2. Glueballs \hfil} 
\bigskip
 \line{\elevenit 2.1. Models for Gluon Fields\hfil}
\smallskip
The low energy hadron spectrum was  well described by the naive quark
model before the advent of QCD.
The gluonic degrees of freedom in QCD are quite a challenge from
this point of view since non-perturbative methods are needed to
discuss the hadron spectrum. Early proposals for hadronic states
built of glue came from dual pomeron arguments$^3$ and from
construction of gluonic hadron operators$^4$. More comprehensive
treatments  have the
requirement  to build a model which encompassed the
confining force between quarks. Here I will discuss two early
 models which gave predictions for glueball states and which
have been a fertile source of intuition$^{5,6}$. These models
build in  confinement explicitly.

One of the earliest QCD-based approaches to hadron spectroscopy was the MIT
bag model$^{7}$. In the cavity approximation, this model
could treat quarks and gluons in a unified way by
putting them in a spherical bubble of empty (or perturbative)
vacuum inside the extensive non-perturbative vacuum.
The lightest gluonic mode in the bag$^5$ is  TE ($J^P=1^{+}$ ) with
the next mode  TM ($J^P=1^{-}$).
The model inevitably had some freedom - basically
the bag constant which describes
the strength of the confining force. If this is assumed to be the
same for quark and gluonic modes then the TE  mode is $34\%$ heavier
than the lightest quark mode. Thus the bag model tends to suggest
relatively light glueball states. However,
this ratio of the energy of gluon to quark modes does not take account of
self energies and these are known to be important in the quark sector.
The gluon self interaction has not been calculated and
thus the bag model has considerable flexibility in absolute
glueball mass predictions.

What does remain are estimates for the ordering of the glueball masses.
 Taking the lowest energy modes of the
gluon field in a spherical bag  and combining them into colour singlets
gives the  glueball spectrum. Neglecting interactions between these
constituent gluons then gives the basic  prediction that
$0^{++}$ and $2^{++}$ states are lightest (from TE plus TE), followed by
$0^{-+}$ and $2^{-+}$ (from TE plus TM).
 Some treatments have also proposed a $1^{-+}$
state which would be of great importance experimentally since it is
a spin-exotic. The debate arises because such a state  can be made out
of two massive $J=1$ gluonic modes but not from two massless modes.
A more field theoretic treatment$^8$ of gluonic operators strongly
supports the absence of this light spin-exotic.

The simple bag model has several defects - perhaps
the most obvious is that the centre of mass motion is incorrect.
Also in the cavity approximation, confinement is built in in a very
 explicit way which limits calculations that can be performed.
An attempt to generalise  the features of such models has been
made$^8$. The basic idea is that hadronic states have a mass related
to the dimension of the lowest dimension operator that can create
them from the vacuum.  Writing $q$ for quark fields, $G$ for gluons,
$\Gamma$ for a generic Dirac matrix and $D$ for a covariant derivative:
the lowest dimension hadronic operators are

\smallskip
\halign{\hskip 0.5in # \hfil & \qquad $#$ \hfil \cr
dim=3 & \bar{q} \Gamma q   \cr
dim=4 & \bar{q} \Gamma D q,\,  GG  \cr
dim=5 & \bar{q} \Gamma DD q, \,  GDG,\,  \bar{q} \Gamma G q \cr
dim=6 & GDDG, \, GGG, etc \cr
}
\smallskip
\noindent For glueballs, this scenario$^{8}$ leads to
$0^{++},2^{++},0^{-+},2^{-+}$ from $GG$ and
then $1^{++},3^{++}$ from the heavier $GDG$. These glueball states
could be expected to lie at energies close to the $P$ and $D$-wave
quark model excitations (ie 1-2 GeV).

An alternative  model which encompasses confinement is the strong coupling
lattice model$^{6,7}$.
In principle the strong coupling model taken to sufficiently high
powers of the inverse coupling can give exact answers -
we will discuss this later. Here I will
discuss low-order strong coupling as a model which provides a
vocabulary for gluonic excitations.
 This approach is explicitly gauge invariant and
treats the gluon field as a flux link on the hypercubic lattice.
Such a flux link can be from quark to antiquark to provide the
usual binding, or it can be a disconnected loop of colour flux. This
latter possibility is the prototype for a glueball.
 For those closed loops of colour flux, the analysis
 of rotational degrees of freedom in lattice
calculations is in terms of the cubic symmetry group $O_h$. The
relationship between representations of $O_h$ and those of the
continuum $SU(2)$ rotations have been worked out$^{9,10}$.
In such a strong coupling picture, the lightest glueball
 is the shortest loop (the
$1 \times 1$ plaquette) and this corresponds to $0^{++}$ , $2^{++}$
and $1^{+-}$.
More complicated shapes allow all $J^{PC}$ - in particular the
so-called spin-exotic values such as $0^{--}, 0^{+-}, 1^{-+}, 2^{+-}, $
etc. These states are not allowed in the quark model, so are a natural
way to discriminate experimentally between glueballs and ordinary
mesons.
 With the realisation that even high order strong coupling
calculations are insufficient to reproduce the asymptotic freedom
behaviour of QCD, emphasis has shifted to lattice simulation. We
discuss in detail that approach in the next sub-section.

A continuum model inspired by the lattice approach is the flux
tube model$^{11}$. This can describe gluonic excitations in both
glueballs (closed flux loops) and hybrid mesons (excited flux lines
between fixed ends). The energy scale of the glueball states
is somewhat flexible whereas tighter predictions apply to
hybrid energy levels. The model as proposed$^{11}$ has low-lying
glueball states consisting
of a $0^{++}$ ground state at
1.5 GeV, then $1^{+-}$ at 2.2 GeV and $2^{++}$ at 2.7 GeV. These
are the same three lowest states as in the low-order strong coupling model.

The models discussed all give the prospect of finding glueball states in the
hadronic spectrum. The properties of such states are that they should
have no flavour quantum numbers and should be additional to quark-antiquark
mesons. This is not a great restriction since orbital and radial
excitations of $q\bar{q}$  are present to confuse interpretation
of the spectrum, and there are also molecular ($q\bar{q}q\bar{q}$)
and hybrid ($q\bar{q}+$glue) possibilities too. Thus from the
early days it was clear that a classification by production mechanism
and/or decay was essential. The glue-sniffers rule-of-thumb is that
preferred production mechanisms are those which are
glue-rich: OZI suppressed processes,
double Pomeron production, etc. Moreover the decays should be
flavour blind so, for example, $K\bar{K}$ and $\eta\eta$
should be more prevalent
than in most quark model decays.

Of course these ideas are really just a mantra - to be repeated often
but without any certainty of implementation. As we shall see below, the
truth may well be that hadronic states are complex mixtures of
gluonic and quark degrees of freedom and no idealised glueball exists
at all.

\bigskip
 \line{\elevenit 2.2. Lattice Simulation of Glueballs\hfil}
\smallskip
To resolve the question of the gluonic
degrees of freedom in hadron spectra, what is needed theoretically
is a quantitative non-perturbative analysis of QCD. The only
comprehensive approach available at present is  provided by lattice
gauge theory. Although high order calculations in the strong-coupling
formalism can provide useful estimates of the glueball spectrum$^{12}$,
the main source of accurate results comes from lattice
simulation in Euclidean space-time.  This lattice approach aims to
derive exact results directly from the QCD lagrangian with no
further model input. As yet full QCD with realistic light quark
masses is not accessible because of computational limitations -
nevertheless sufficient information is available to give very strong
guidance.

 A possible alternative method is provided
by QCD sum rules$^{13}$ in which non-perturbative features are cast into
vacuum condensates which can be determined in principle from experiment.
Such an approach is less comprehensive in that some spectroscopic input
 is usually needed
to get further output. Also there does seem to be a subjective
element in such sum rules - different authors can get rather different
conclusions.

Returning to lattice simulation: there are many cross-checks which
need to be made before results can be considered as applicable to the
real continuum, large-volume world. These are that the results
(eg dimensionless ratios of masses) are independent of the lattice
spacing $a$ once $a$ is small enough, that the lattice size $L$ is
large enough, and that rotational (and Lorentz) invariance are
restored.  These checks have been established satisfactorily for
quenched (or pure-gauge) QCD. This is the theory with no dynamical
quarks - so the lagrangian contains just the gluon self-interaction. This
is already a non-trivial theory which exhibits confinement.
Indeed the vacuum shows a behaviour very far from the empty perturbative
vacuum - instead there are whirlpools of colour flux which provide
a rather disordered  medium.

 But,
as ever, the theorists have responded by giving the correct answer
to the wrong question!  The effect of quark-antiquark pairs on the
quenched vacuum still needs to be taken into account. Attempts to do
that require rather massive quarks at present to keep the
computational requirements within those currently available.
With such massive dynamical quarks the net effect is rather small.
In general one would still hope to learn about the
relative importance of dynamical quark loops from such studies.
However, extrapolating such
effects from heavy dynamical quarks may be misguided because
some important contributions may have been omitted.  For example
the $\rho$ is above the $\pi\pi$ threshold for $P$-wave angular momentum
in all simulations so far and thus no contribution from
$\rho$ decay is present in the simulation but will surely set in for
lighter quark masses.
 So it is possible that, for some processes,
quite new mechanisms may set in rather
abruptly as the quark masses are reduced. However, in other examples
such as the $\beta$-function it is known from perturbation theory
that the effect of the light quark loops is only about 20\% of the
gluonic component. So the quenched approximation may still be
a good guide in many cases.

The glueballs are of course only clearly defined in the quenched
approximation. As soon as quark loops are allowed then mixing can
occur between gluonic states and quark-antiquark states. So it
is important to establish the glueball spectrum in the quenched
approximation. This has been explored by lattice simulation for
over a decade. Early work suffered from lattice spacing too large
and  lattice volume too small. With the computational resources
now available these limitations can be overcome. In fact lattice
results have been stable for the last few years. The most
comprehensive study gives results$^{14}$:

\smallskip
\halign{\hskip 0.5in $#$ \hfil & \qquad # \hfil \cr
0^{++} & 1.54(9) GeV \cr
2^{++} & 2.33(9) GeV \cr
0^{-+} & 2.77(31) GeV \cr
0^{++} & 2.77(18) GeV \cr
2^{-+} & 2.90(44) GeV \cr
1^{+-} & 3.04(22) GeV \cr
}
\smallskip

\noindent For the states shown  above, the lattice
results are consistent for different lattice spacings, different
volumes, and different shapes of gluonic operators (for example
the 2-dimensional $E^{++}$ and 3-dimensional $T_2^{++}$ representations
of $O_h$ are found to form the 5-dimensional $2^{++}$ level).

The lattice method involves measuring correlations of operators
with the correct symmetry at separation $T$. This gives upper
limits to the mass values for a given quantum number since an
admixture of excited and ground state can contribute. To isolate
the ground state one needs to use several different operators (formed
from different shapes and different smearing levels). Furthermore
by comparing results as $T$ is increased one can check stability
since the relative contribution of the ground state is enhanced as
$\exp(\Delta E \, T)$ where the next state has energy $\Delta E$ higher.
This situation is represented in  fig. 1  where the effective
energy value coming from the $T$-ratio
2/1 is shown with  the upper errors
representing the statistical errors while the lower error bars
represent an estimate of the  systematic error coming from the
difference of 1/0 and 2/1 estimates.
For the heavier states shown, provided that the
 flux-loop operators chosen are an
adequate basis for the ground state of that representation,
then the upper bound should be close to the actual value.
  Some $J^{PC}$ values are not illustrated since
no evidence existed that a reasonable basis had been found.
Thus fig.~1 gives reasonable estimates of the glueball spectrum
ordering for most $J^{PC}$ values.

\topinsert
\vskip 6in
{\tenrm  \noindent \baselineskip=12pt
Figure 1.
 The glueball spectrum from lattice gauge theory
simulation$^{14}$. Results are quoted as ratios to the $0^{++}$
ground state and are also given in GeV obtained by setting the
string tension $\sqrt K=0.44$ GeV. The upper error bars are
statistical and the lower error bars are estimates of the
possible systematic error.}
\endinsert

The lattice calculations yield
dimensionless ratios - conventionally in terms of the string
tension $K$ which is  determined relatively precisely in lattice
simulation. Expressing the results in terms of GeV, one needs
to relate to a physical measurable and again the string tension
is a natural candidate with a value $\sqrt K=0.44 $GeV obtained from
charmonium and $b\bar{b}$ spectroscopy and from the slope
of Regge trajectories. Indeed this value has been used
to construct the table given above. This value, however, is from the world
of experiment with the full dynamical light quark vacuum. Thus there
 could easily be systematic errors of 20\% or so in this energy
assignment. This is quite important for the interpretation of
the glueball results because it allows an overall mass shift
by this unknown factor.

Coming back to the lattice results, we see  that the lightest
glueball is a $0^{++}$ at around 1.5 GeV with next a $2^{++}$ at
around 2.3 GeV and then several states near 2.8 GeV ($0^{-+},
2^{-+}, 1^{+-}$ and excited $0^{++}$).
This level ordering can be compared with the descriptive models
discussed previously. The ordering by the dimension of the
hadronic operator  suggests
$0^{++}$ and $2^{++}$ lightest, then $0^{-+}$ and $2^{-+}$ with
$1^{++}$ and $3^{++}$ next. The lattice results agree roughly for
 the four lowest states but the $1^{+-}$ is relatively light and this
would have to come from a $GGG$ operator of dimension $6$. The lattice
low-order strong-coupling has low-lying $0^{++}, 2^{++} $ and $1^{+-}$ which
is again a rough guide. Thus for the lowest states,
the energy ordering observed is rather similar to that in the simple models
discussed above. However, neither model is particularly accurate in
giving either the ordering or the absolute energy scale.

It is of interest to compare the quenched lattice glueball spectrum
with Pomeron phenomenology. The idea being that glueball states
can occur if a Pomeron Regge trajectory crossed an integer $J$.
 The phenomenological trajectory$^{15}$ of $1.085+0.25m^2$ then yields
a $2^{++}$ state at 1.9 GeV
which is not very far from the lattice result quoted above.

 There is no sign of any
spin-exotic glueball state (marked E in fig. 1)
at an energy below twice the $0^{++}$
 ground state. Claims for such a low energy $1^{-+}$
 signal in lattice calculations
a few years ago were most likely due to misinterpretation
of the $2^{-+}$. This lack of a low-lying exotic is a pity
 - such an exotic state could not
mix with a $q\bar{q}$ state so would have been
a good candidate  to identify experimentally. Fig. 1 shows
some evidence for a $1^{-+}$ exotic near $2.8 \, m(0^{++})$. There is also
evidence$^{16}$ from lattice simulation of $SU(2)$ colour that an exotic
$1^{-}$ state exists with mass about 2.8(5) times the $0^+$
glueball. Since in other respects the $SU(2)$ and $SU(3)$ results
are extremely similar this strengthens the suggestion
 that a $1^{-+}$ state may indeed
be present at such a mass. Even so, at a mass as high as this
(circa 4 GeV) there is little prospect of such a particle being
observed as a narrow, unmixed glueball.

As we have discussed, dynamical quark calculations on a lattice are
still rather exploratory. The glueball spectrum has been studied$^{17}$ but
with quark masses several times too large. As a result no
significant departures are seen from the quenched result. This, as
we have emphasised, does not mean that significant effects will
not occur as the quark masses are reduced to the physical values.

Lattice simulation can also give information on wavefunctions as well
as on mass spectra. For glueballs one way to characterise their
spatial extent is to look at the energy-momentum distribution.
Exploratory work$^{18}$ shows a distribution with a radius
extending to $4m^{-1}$. From a study of the operators used to create
a glueball from the vacuum, it is also possible to gain information
about the size but this information is harder to interpret.

\bigskip
 \line{\elevenit 2.3. Experiment and Glueballs\hfil}
\smallskip
{}From these lattice results, we can attempt to identify experimentally
observed hadrons as glueballs. One of the prime candidates is the
G(1590) which has been observed$^{19}$ by the GAMS group in
peripheral and central production. It is consistent with a scalar,
has a width of 175 MeV  and has
prominent decays to $\eta\eta$ and $\eta\eta'$. These decay modes
suggest disconnected quark diagrams (ie OZI suppression) since the
$\eta$ and $\eta'$ have substantial $s$ quark components. Indeed
a glueball decay would necessarily proceed by disconnected diagrams.
To be more quantitative,
 the $\eta'/\eta$ ratio observed for G decay of 3/1 is consistent
with the ratio observed for the disconnected processes
 $J/\psi \rightarrow \eta + \gamma$ to $J/\psi \rightarrow \eta' +
\gamma$.  The production of the G is relatively stronger
in central than peripheral processes which is again evidence for
a gluonic coupling.  A clinching argument would be the observation
of production of the  G
in $J/\psi \rightarrow G + \gamma$. No such signal is reported but
the estimate of the branching ratio from QCD sum rules$^{20}$ is
sufficiently small for the signal to have escaped detection.
Recently the G has been observed by
an  experimental group$^{21}$ at LEAR in $p\bar{p} \rightarrow
\pi^0 \eta \eta $ as a $0^{++}$ state at 1560(25) MeV with a width
of 245(50) MeV. This confirms the existence of the resonance and
supports its position as a candidate for
a hadron with non-quark-model constitution.
Thus the G(1590) is consistent with being a glueball and, moreover,
its mass is in the region expected from quenched lattice calculations.

One of the sources of glueball candidates is indeed in the reaction
$J/\psi \rightarrow \gamma + ...$. Here one has a disconnected quark
diagram since the $c\bar{c}$ must annihilate. Thus glueball
production should be relatively enhanced (once disconnected) compared
to quark-antiquark meson production (twice disconnected). The $i(1450)$ and
$\theta(1700)$ signals have been much discussed as glueballs. It is,
however, now rather clear that both these regions are complex with
several hadronic states so that interpretation is confused. This is
consistent with the lattice results which would favour considerably
higher masses for glueballs.

Another source of experimental signals which is  often discussed
is the pair production of OZI violating states. A
typical example is $\phi \phi$
production by non-strange processes. The difficulty in confirming
a glueball assignment in these cases is that the observed structures
are close to threshold so that a molecular interpretation is
also viable. Indeed there do seem to be extra states attached to
thresholds which are molecular - the deuteron is the most clear-cut
example.

Because mixing between glueballs and quark-antiquark states is
allowed, it is unclear whether any clear-cut glueball state should
exist. One way to explore this would be theoretically: varying the
light quark masses in full QCD lattice simulation from heavy
(effectively the quenched approximation with well defined glueballs)
to light (the physical spectrum) to map out the relationship
between glueballs and the physical hadrons. This programme is as yet
not feasible because of computational limitations. In the absence
of such guidance, one has to rely on rather subjective estimates
of the status of experimental glueball candidates. Indeed there has
been little progress in establishing such a candidate in the last
decade.
But of course, the experimental detection of a glueball is always likely
to be ambiguous, since a glueball is mainly defined by what it is
{\elevenit not} rather than what it is!

\bigskip
 \line{\elevenit 2.4. Gluelumps\hfil}
\smallskip
An intriguing possibility which would give much insight into gluonic
contributions to hadronic spectra is provided by the
 so-called gluelump$^{23}$.
This is a hadron consisting of a static colour octet. In other
words it is like a glueball with one gluon nailed down. Such a state
could occur if a massive colour octet state existed - such as
a gluino. Then the gluino, if relatively stable, would attract a gluonic
field to make a colour singlet hadron - the glueballino. This
possibility has been discussed in the bag model$^{22}$. Also lattice
simulation in the quenched approximation has been used$^{23}$ to
analyse the spectrum and  to study the distribution of the
wave function of the gluonic field. The lattice results$^{23}$ are
most comprehensive for $SU(2)$ colour and they give a
gluon field of $J^{P}=1^{+}$ as the ground state with a $1^{-}$ state
some 200 MeV above it. This ordering is the same as that of the bag
model in the cavity approximation although the detailed distribution
of colour electric and magnetic fields is different. The lattice
results suggest a diameter of order 1 fm for the gluelump.

\vskip 0.6cm
\line{\elevenbf 3. Small Volume QCD: the Femto Universe\hfil}
\vglue 0.4cm
The theoretical world of a small spatial volume of size $L$ with periodic
boundary conditions has several nice features. The most important is
that perturbative techniques can be used to study the hadron spectrum.
This arises because the minimum momentum $p=2\pi/L$ and so may be
large enough at small $L$ for all the modes with non-zero momentum
to be treated perturbatively since the effective running coupling
for those modes will be small at high momentum. Thus one can integrate
out the gaussian fluctuations of the non-zero momentum modes and
derive an effective lagrangian in the remaining homogeneous (ie zero
momentum) mode$^{24}$. The resulting effective lagrangian in the
 zero momentum modes can be extended to apply at intermediate volumes by
taking into account tunnelling between different vacua$^{25}$.
The spectrum has been studied most comprehensively for
$SU(2)$ gluonic colour fields. The spectrum is obtained either
 by solving an effective
hamiltonian by variational techniques$^{25}$ or equivalently the effective
lagrangian can be tackled by simulation$^{26}$.
 These semi-analytic results can be compared with conventional lattice
simulation in a small periodic volume$^{27}$. Excellent agreement
is found. This is very encouraging and indicates that we have a reliable
knowledge of the small-volume spectrum.

As well as purely gluonic fields in a small spatial volume,
one can include light quarks too. With anti-periodic boundary
conditions for the fermionic quarks, they have non-zero momentum
and so can be integrated out. The resulting effective lagrangian
can be investigated to give the spectrum$^{26}$. One surprise
is that the glueball states (defined as flux loops that don't
encircle the periodic spatial boundary) and torelon states (defined as
flux loops that do encircle the periodic spatial boundary) can mix
as soon as quarks are included and this mixing gives very dramatic
effects. This should serve as a warning that dynamical-quark lattice
calculations will have to use large spatial sizes to avoid this
mixing by making the torelon states very heavy.

Attempts to extend the intermediate volume approach to larger
volumes have been unsuccessful. For $L$-values greater than $5/m$,
where $m$ is the lightest glueball mass, significant new
dynamical effects appear. The non-zero momentum modes become non-trivial,
 instanton effects appear, and  the effective lagrangian approach
looses its simplicity. This conclusion also indicates that lattice
results need to use a sufficiently large spatial volume to avoid
the substantial finite size effects that set in for $L < 5/m$.
Indeed quenched lattice simulations indicate$^{14}$ that $L > 9/m$ is
needed to avoid finite-size effects.

\vskip 0.6cm
\line{\elevenbf 4. Hybrid Mesons \hfil}
\vglue 0.4cm
We define a hybrid meson as a $q\bar{q}$ system with additional
gluonic excitation. Some authors had proposed ``hermaphrodites'' $^{28}$
and ``meiktons'' $^{22}$ for such systems but the title hybrid has
become accepted as a compromise.
The definition of a  hybrid meson is less clear than for
a glueball since even the basic
quark model mesons have a gluonic component which is responsible
for the binding force. So we must establish that the gluonic
component is excited before labelling a state as a hybrid meson.
In the bag model approach$^{5,28,22}$,
 an extra gluonic mode is added - amounting
to a constituent gluon of $J^P=1^+$ in the TE mode
or at a slightly higher energy $J^P=1^-$. This ordering of
 gluonic modes agrees
with those found around a static adjoint colour source in lattice
simulation so is a reliable guide.
Then the lowest lying hybrid mesons should be obtained by combining
this gluon mode with the $q\bar{q}$ lowest mode. This gives $J^{PC}$
values of $1^{--}, 0^{-+}, 1^{-+}, 2^{-+}$. Here the main interest
is the $1^{-+}$ state which is spin-exotic. The bag model gives a
dense spectrum of hybrids at higher energies: since gluonic, radial
and orbital excitations are available. However, there is
 no very convincing reason for most of
these $q\bar{q}G$ states to remain unmixed with $q\bar{q}$ mesons.
Moreover, in view of the lack of detailed agreement between bag model
and lattice glueball spectra, it would be useful
 to find  alternative descriptions of gluonic excitations in hybrid mesons.

This is provided by  lattice gauge theory which describes the
gluon field by colour flux along the links of the lattice. Then one can
visualise the ground state mesons in the quark model as being made
of a quark and antiquark joined by a string of this colour flux. In the
ground state the string will be as short as possible - and symmetric
about the interquark axis. This leads naturally to the possibility of
states with the string excited: either to larger transverse fluctuations
or to less symmetric states with respect to the interquark axis. One
feature of this approach is that these string excitations can carry
non-zero angular momentum about the interquark axis. This angular
momentum then combines with rotational and spin angular momentum to
give the resulting $J^{PC}$. These have been worked out$^{29}$ and, indeed,
one consequence is that spin exotic combinations are allowed.

\topinsert
\vskip 6in
{\tenrm
\baselineskip=12pt
\noindent
 Figure 2. The energy levels of the potential
between static quark and antiquark at separation $R$ from
lattice gauge theory simulation$^{30}$.
The labels $A_{1g}$ and $E_u$ refer to ground state and first
excited gluonic fields respectively. The dotted curve reproduces
the experimental $\Upsilon$ spectrum and has a Coulomb strength
enhanced compared to the lattice potential as described in the
text.  The spectrum of mesons
in these potentials for the $b\bar{b}$ system is illustrated.
We illustrate only the lowest hybrid level - note that this  contains
exotic $J^{PC}$ values.}
\endinsert

The continuum version of the lattice approach is provided by the flux tube
model$^{11}$.
This model has been used to give predictions for hybrid
mesons containing light quarks. These levels are expected to be
relatively heavy: above 1.8 GeV. This energy value comes basically
from the string model which gives excited  levels with energies
$n\pi/R$ higher for integer $n$. Then $R$ which is the
interquark separation can be estimated from
standard quark models.  Moreover there is again a rather
dense spectrum with eight degenerate nonets expected. These are
$0^{\pm \mp}, 1^{\pm \mp }, 2^{\pm \mp}, 1^{\pm \pm}$ which again
include spin-exotic states.

Now it remains to evaluate the energy associated with these excitations
of the interquark colour flux. A relativistic string with fixed ends
of length $R$ has excitations $n\pi/R$ for integer $n$. The
gluonic flux between heavy quark and antiquark is not necessarily
well described as a relativistic string of zero intrinsic width. This
can be explored by lattice simulation of the interquark potential and
its excitations. The work of the Liverpool group$^{30}$ has determined
this for the quenched approximation. Their results are that for large
$R$, the simple string estimate is a good guide but that at smaller $R$
there are some detailed changes. The first excited potential lies in
the $E_u$ representation of the symmetry group $D_{4h}$ and corresponds
to an angular momentum component of one unit about the interquark axis.
As shown in the fig.~2, the corresponding potential is rather flat at
smaller $R$.
The excited potentials are found to be stable in the quenched approximation:
that is they cannot decay by glueball emission to the ground state
potential. It is useful to compare the lattice potentials with
the string model.
It turns out  that a version of the string model with fixed
ends at separation $R$ gives potentials
$$
 V_{n}(R) = { \left ( K^2 R^2 - \pi/6 + 2 \pi n K \right ) } ^ {1/2}
$$
which yields $V_n(R)-V_0(R)$ behaving as $n\pi/R$ at large $R$
but is a better description of the lattice results
at intermediate $R$-values. Here $K$ is the string tension. This
expression is illustrated for $n=1$ in fig. 2 where it is seen
to describe the $E_u$ excitation well for $R > 2$ GeV$^{-1}$.

These lattice results for the excited potentials between static colour
sources
can be directly applied to hybrid mesons containing heavy quarks. For
the $b\bar{b}$ system in particular, a non-relativistic potential
 description is accurate. Furthermore the adiabatic approximation
can be applied: namely that the gluon field adapts instantly to the
quark and antiquark as their axis rotates. This approximation can be
justified if
the energy of the gluonic excitation is large compared to the quark
radial and orbital excitations and we find this satisfied.
So a Schr\"odinger equation treatment gives the hybrid meson spectrum.
 The first excited potential lies in
the $E_u$ representation of the symmetry group $D_{4h}$
which results in hybrid states with $J^{PC}$ values of
$0^{\pm \mp}, 1^{\pm \mp }, 2^{\pm \mp}, 1^{\pm \pm}$
which includes spin-exotic values. For the $b\bar{b}$ system the
lightest such hybrids are estimated$^{30}$ to lie at
 1.11(3) GeV above the $\Upsilon$ while for $c\bar{c}$ they are
at 0.94(3) GeV above the $J/\psi$..  These quenched lattice
results$^{30}$ for the excitation spectrum in these  central potentials
are collected here for the low-lying $b\bar{b}$ system
using $m_{b}=$4.64 GeV and $\sqrt K =$ 0.44 GeV:

\vbox{
\smallskip
\halign{\hskip 0.5in $#$ \hfil & \qquad # \hfil & \qquad # \hfil \cr
   & $\Delta E_{quenched}$ & $ \Delta E_{phen}$ \cr
b\bar{b}:  & & \cr
 0^{-+}, 1^{--} & 0.00 GeV & 0.00 GeV \cr
1^{+-}, 0^{++}, 1^{++}, 2^{++}  & 0.32 GeV & 0.45 GeV \cr
2^{-+}, 1^{--}, 2^{--}, 3^{--}  & 0.45 GeV & 0.56 GeV \cr
hybrid:& & \cr
0^{\pm \mp}, 1^{\pm \mp }, 2^{\pm \mp}, 1^{\pm \pm} &
 1.11 GeV & 1.36 GeV \cr
}
\smallskip
}

This evidence, however, must be taken with some caution since the
quenched
lattice ground state potential does not reproduce the upsilon levels
precisely. This can be seen from the above table where the ratio
$ (1P-1S)/(2S-1S)$ is $0.7$ rather than $0.8$.
 This can be understood as coming partly from the different
one gluon exchange strength with dynamical quarks rather than purely glue -
because of the $33-2N_f$ factor in the denominator of the running
coupling. This gives a 22\% increase in the one gluon exchange
 coupling strength - whereas an increase of the quenched lattice value
by more like 70\% is needed to fit the spectrum.
Another possible modification to the quenched results
is that the large $R$
potential will saturate due to string-breaking from $q\bar{q}$ pair
creation.
To estimate the size of possible effects arising from these changes
 to the ground state potential, the lattice potentials can be modified
by an enhanced one gluon exchange
component so that they fit the $b\bar{b}$ spectrum. The results
with such a phenomenological potential (with the $A_{1g}$ potential
modified by increasing the coefficient of the $1/R$ term) are also
shown in the table above.  So the
lightest  hybrids are estimated$^{30}$ to lie at
 1.36 GeV above the $\Upsilon$ while for $c\bar{c}$ they are
at 1.07 GeV above the $J/\psi$. Thus there is a systematic
error of about 0.25 GeV on the quenched hybrid levels.
Hence, although some of the dynamical quark effect appears to
go the wrong way,  there is sufficient uncertainty that it is possible
for the lightest hybrid level to end up below threshold (the
 thresholds,
relative to the $1S$ levels, are at
1.10 and 0.63 GeV for $B\bar{B}$ and $D\bar{D}$ respectively)
- and so be more  easily detectable experimentally. Note that
the $b\bar{b}$ hybrids are expected to lie closest to the open
threshold and so are the most promising place to look for
narrow hybrid resonances.

Because the $E_u$ potential is rather flat, the wave function will
be extended and there will be many hybrid states rather close
in energy since radial and orbital excitations in such a flat
potential are of low energy.  There will be many different $J^{PC}$
values, among them exotic ones. This exotic spin-parity is the
most promising experimental signature. The decay width of these
states will be quite narrow if they lie below the two meson threshold
(ie $B\bar{B}$) but otherwise rather wide since no quantum numbers or
dynamical factors inhibit decay to two such mesons.

As well as exotic spin-parity states, another signal for hybrid mesons
may be additional states in the vector channel which can be probed
directly from $e^+e^-$. Because of the extended wave function, the
wave function at the origin will be small for such hybrids and
hence they will couple weakly to $e^+e^-$. It may none-the-less be
possible to detect their presence in
 detailed $B$-factory studies.

All the above comments concerning the $b\bar{b}$ system apply equally
well
to charmonium of course, with the proviso that non-relativistic
potential models are somewhat less reliable here.  Turning to light
quarks, however, is a very different matter. The potential approach
has no reason to apply any more. A direct lattice investigation of the
light quark hadron spectrum is feasible of course. Indeed it has been
carried out extensively. The hybrid states, however, are likely to be
relatively heavy and are thus difficult to extract numerically. While
every effort should be made to look for spin-exotic light quark
mesons, it may be difficult to obtain convincing signals from
lattice simulation. Certainly non-local operators will be needed to
create these states from the vacuum - so that a gluonic component
is present.

Experimentally, the spin-exotic signal is a preferred indication of
the possibility of a hybrid meson assignment. One such state which is
worth discussing is the $1^{-+} \ M(1406)$ which is seen$^{31}$
decaying into $\eta \pi$ in a P-wave, so with $I=1$.
 The main drawback to an
interpretation as a hybrid is the relatively low mass. Most theoretical
schemes involve a rather dense spectrum of hybrid levels and these
are not seen around such an energy value. An alternative
 interpretation as a $qq\bar{q} \bar{q}$ state is viable$^{1}$.
Thus yet again a clear interpretation of experimental signals is
lacking.

\vskip 0.6cm
\line{\elevenbf 5. Conclusions \hfil}
\vglue 0.4cm

The energy levels associated with gluonic excitations are well
established in the quenched approximation to QCD. These energies
are of the order of $1$ GeV. This value is large compared
to the quark model orbital and radial excitation energies  which
explains the comparative success of the quark model.
A full theoretical understanding of the gluonic component of
hadronic states can really only come from a comprehensive
non-perturbative determination of the QCD spectrum. At present, the only
prospect for such an analysis is from lattice simulation. With
present algorithms, a considerable increase in computing power is needed
for such a full study. The result, however, would be very valuable
and interesting. One could study theoretically the spectrum as the
light quark masses were varied from their physical values to
heavy values where the quenched approximation is valid and we already
know the spectrum. This would settle unambiguously the continuing
controversy in trying to assign gluonic components to the physical
hadrons.

\vskip 0.6cm
\line{\elevenbf 6. References \hfil}
\vglue 0.4cm
\medskip
\item{1.} F. E. Close, {\elevenit Rep. Prog. Phys.}
{\elevenbf 51} (1988) 833.
\item{2.} T. H. Burnett and S. R. Sharpe, {\elevenit Ann. Rev. Nucl.
Part, Sci} {\elevenbf 40} (1991) 327.
\item{3.} P. G. O. Freund and Y. Nambu, {\elevenit Phys. Rev. Lett.}
 {\elevenbf 34} (1975) 1646.
\item{4.} H. Fritzch and P. Minkowski, {\elevenit Nuovo Cimento}
 {\elevenbf A 30} (1975) 393.
\item{5.} R. L. Jaffe and K. Johnson, {\elevenit Phys. Lett.}
 {\elevenbf B 60} (1976) 201.
\item{6.} J. Kogut, D. K. Sinclair and L. Susskind, {\elevenit Nucl. Phys.}
 {\elevenbf B 114} (1976) 114.
\item{7.} T. De Grand, R. L. Jaffe, K. Johnson and J. Kiskis,
 {\elevenit Phys. Rev.} {\elevenbf D 12} (1975) 2060.
\item{8.} R. L. Jaffe, K. Johnson and Z. Ryzak,
 {\elevenit Ann. Phys.} {\elevenbf 168} (1986) 344.
\item{9.} B. Berg and A. Billoire,
 {\elevenit Nucl. Phys.} {\elevenbf B 221} (1983) 109.
\item{10.} C. Michael,
 {\elevenit Acta. Phys. Polonica} {\elevenbf B 21} (1990) 119.
\item{11.} N. Isgur and J. Paton
 {\elevenit Phys. Lett.} {\elevenbf B 124} (1983) 247.
\item{12.} C. Hamer,
 {\elevenit Phys. Lett.} {\elevenbf B 224} (1989) 339.
\item{13.} M. Shifman et al.,
 {\elevenit Nucl. Phys.} {\elevenbf B 147} (1979) 385,448;\hfil \break
 V. Nokikov et al.,
 {\elevenit Nucl. Phys.} {\elevenbf B 165} (1980) 55.
\item{14.} C. Michael and M. Teper,
 {\elevenit Nucl. Phys.} {\elevenbf B 314} (1989) 347.
\item{15.} P. V. Landshoff,
 {\elevenit  Proceedings of Aachen QCD Workshop}.
\item{16.} S. J. Perantonis and C. Michael,
 {\elevenit Nucl. Phys.(Proc. Suppl.)} {\elevenbf B 20} (1991) 177;
 {\elevenit J. Phys.} {\elevenbf G } in press.
\item{17.} K. M. Bitar et al.,
 {\elevenit Phys. Rev.} {\elevenbf D 44} (1991) 2090.
\item{18.} G. A. Tickle and C. Michael,
 {\elevenit Nucl. Phys.} {\elevenbf B 333} (1990) 593.
\item{19.} F. G. Binon et al.,
 {\elevenit Nuovo Cimento} {\elevenbf 78 A} (1983) 313;
                           {\elevenit 80 A} (1984) 363;\hfil \break
 D. Alde et al.,
 {\elevenit Nucl. Phys.} {\elevenbf B 269} (1986) 485;
 {\elevenit Phys. Lett.} {\elevenbf B 201} (1988) 160.
\item{20.} S. Narison and G. Veneziano,
 {\elevenit Int. J. Mod. Phys.} {\elevenbf A 4} (1989) 2751.
\item{21.} C. Amsler et al.,
 {\elevenit CERN preprint} PPE-92-114.
\item{22.} M. S. Chanowitz and S. R. Sharpe,
 {\elevenit Phys. Lett.} {\elevenbf B 126} (1983) 225.
\item{23.} I. H. Jorysz and C. Michael,
 {\elevenit Nucl. Phys.} {\elevenbf B 302} (1987) 448.
\item{24.} M. L\"uscher,
 {\elevenit  Nucl. Phys.} {\elevenbf B 219} (1983) 233.
\item{25.} J. Koller and P. van Baal,
 {\elevenit  Nucl. Phys.} {\elevenbf B 302} (1988) 1.
\item{26.} J. Kripfganz and C. Michael,
 {\elevenit Nucl. Phys.} {\elevenbf B 314} (1989) 25; \hfil \break
      C. Michael,
 {\elevenit Nucl. Phys.} {\elevenbf B 329} (1990) 225.
\item{27.} C. Michael, G. A. Tickle and M. Teper,
 {\elevenit  Phys. Lett.} {\elevenbf B 207} (1988) 313.
\item{28.} T. Barnes and F. E. Close,
 {\elevenit Phys. Lett.} {\elevenbf B 116} (1983) 365.
\item{29.} L. A. Griffiths, P. E. L. Rakow  and C. Michael,
 {\elevenit Phys. Lett.} {\elevenbf B 129} (1983) 351.
\item{30.} S. J. Perantonis and C. Michael,
 {\elevenit Nucl. Phys.} {\elevenbf B 347} (1990) 854.
\item{31.} D. Alde et al.,
 {\elevenit Phys. Lett.} {\elevenbf B 205} (1988) 397.
\bye

\end